# A new approach for uncovering student resources with multiple-choice questions

Nolan K. Weinlader, Eric Kuo, Benjamin M. Rottman, and Timothy J. Nokes-Malach
*Learning Research & Development Center, University of Pittsburgh, 3939 O'Hara St., Pittsburgh, PA, 15224*

The traditional approach to studying student understanding presents a question and uses the student's answer to make inferences about their knowledge. However, this method doesn't capture the range of possible alternative ideas available to students. We use a new approach, asking students to generate a plausible explanation for every choice of a multiple-choice question, to capture a range of explanations that students can generate in answering physics questions. Asking 16 students to provide explanations in this way revealed alternative possibilities for student thinking that would not have been captured if they only provided one solution. The findings show two ways these alternatives can be productive for learning physics: (i) even students who ultimately chose the wrong answer could often generate the correct explanation and (ii) many incorrect explanations contained elements of correct physical reasoning. We discuss the instructional implications of this multiple-choice questioning approach and of students' alternative ideas.

## I. INTRODUCTION

For examining students' conceptual knowledge, qualitative questions, rather than quantitative questions designed for calculations, are the typical assessment tool. Multiple-choice questions infer students' conceptual knowledge from their selected choice; free-response questions might more directly ask students to provide the reasoning behind their response. The use of such conceptual knowledge assessments has made a significant impact in Physics Education Research, uncovering the common difficulties students demonstrate even after instruction [1].

One instructionally relevant question is what these common incorrect answers imply about students' conceptual knowledge? One perspective, consistent with a (mis)conceptions view [2], is that students' incorrect responses reflect their incorrect conceptual knowledge. Instruction, then, should aim to displace this incorrect knowledge and replace it with correct knowledge. By contrast, a resources framework [3] (and a knowledge-in-pieces perspective, in general [4,5]) models students as possessing many cognitive resources, reflecting different ideas for learning and doing physics. With these different resources, students can construct multiple explanations or predictions for a physical situation. Therefore, while a student's final response to a physics question might be the dominant one, alternative explanations may be readily generated from other resources. Supporting this interpretation, prior research has shown that students' initial explanations can shift to alternative explanations over relatively short episodes [6,7].

The resources framework frames the goal of physics instruction as helping students use the right resources in the right ways at the right times, in the face of multiple possible explanations one might generate. When a student's given explanation is incorrect, knowing the alternative explanations that they can generate is especially useful for instruction.

Is the correct explanation in the set of possible alternatives that students' see? One possibility is that the correct explanation will not be in their considered set of alternatives. In this case, instruction may critically need to present the correct explanation as one to consider and promote in one's conceptual reasoning. Another possibility is that the correct explanation is one of the alternatives that students can generate without much difficulty. In this case, the focus of instruction should not be simply to provide the correct explanation but rather to help students reliably choose this explanation in the face of plausible alternatives.

The weakness of typical conceptual knowledge assessments is that they can reveal students' dominant explanations without capturing the alternative ideas available to them. Capturing these alternative ideas can better characterize the state of students' conceptual knowledge and illuminate productive paths forward for instruction. This paper takes on the methodological challenge of uncovering the strongest alternatives to students' dominant reasoning. To do so, we asked students to generate explanations for *all choices* of a multiple-choice physics question. Doing so explicitly prompts students to reveal the alternative explanations that can be constructed without additional instruction. We also ask students to rate how likely they believe that each choice is correct. As opposed to simply choosing which answer they believed was correct, these certainty ratings provided a more fine-grained measurement of the strengths of their conceptual views.

## II. METHOD

### A. Interview Protocol

Sixteen undergraduates who were enrolled in or had taken college physics were interviewed. The interviews were about one-hour long. On a computer, students were led through a sequence of prompts, asking them to make selections on the screen or to provide verbal explanations to a researcher who was observing. Interviews were video recorded to capture these verbal explanations.

Students answered the prompts for 3-6 multiple-choice questions, depending on how quickly they progressed through the interview. In this paper, we discuss results from the first two multiple-choice questions, which all students completed. The Two Boats Q1 and Q2, taken from smartPhysics [8], considered a battleship launching shells at two targets (Fig. 1). These two questions target student understanding of the connection between peak height and time in the air. In Q1, target 2 is hit first, because it has a lower peak height. In Q2, the targets are hit at the same time, because both shells reach the same peak height.

Rather than just having students answer the Two Boats questions, the interview protocol led students through an identical three-part sequence for each question, as follows:

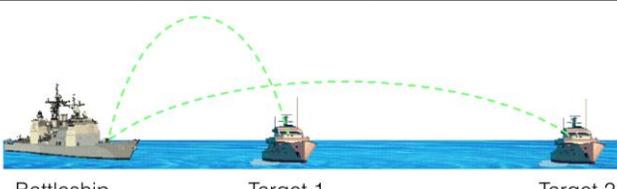
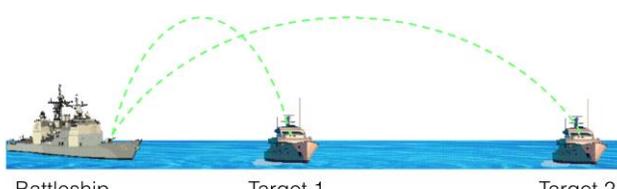

FIG 1. Two qualitative questions asking students to determine the relative flight times of two projectiles.

Part 1 (Answer rating #1) – Students started by rating the likelihood of the answers:

*Here are three possible answer to this question (A, B, C). Rate the likelihood that you believe each answer is correct. 0 – it's not likely at all. 100 – it's definitely correct.*

*A) Target 1 is hit first.*
*B) Target 2 is hit first.*
*C) The ships are hit at the same time.*

Using sliders, students were able to choose whole numbers between 0 (labeled "least likely) and 100 (labeled "most likely") for each multiple-choice option. The total rating for all three had to sum to 100%, as we aimed to capture their belief distributions on a probability scale.

Part 2 (Generate Explanations) – Students were then directed to generate potential explanations for each of the three multiple-choice options:

*For each choice, provide a reasonable explanation for why someone might choose it. Even if you yourself don't think a choice is correct, try to come up with the most convincing explanation that someone who selects that choice might believe (even if you don't believe that explanation yourself!). Please talk out loud about your explanations.*

At the start of the interview, students were told that, when explaining their thinking to the researcher, they might be asked follow-up questions. Here, the follow-up questions aimed to make sure students provided an explanation for all three options and that the researcher understood each explanation.

Part 3 (Answer rating #2) – Students again rated how likely they thought each answer was correct, from 0 to 100. The results reported use this rating as a measure of how much students believed the explanations generated in part 2. The ratings in part 1 are not suitable for this purpose, because it is possible that students have not considered alternative explanations before they explicitly generate them in part 2.

### B. Explanation coding

For Two Boats Q1 and Q2, students' verbal explanations for the three choices, A, B, and C, were coded into different categories. The first and second author generated an initial list of codes by listening to an initial subset of students' explanations. Then, the two coders independently coded all students' A, B, and C explanations for the two questions, discussing disagreements until all codes were agreed upon.

### III. RESULTS

Broadly, the prompt to provide an explanation for all multiple-choice options was successful at uncovering alternative explanations: only one student failed to provide explanations for all three options on one question.

The number of coded explanations is shown in Table 1. These 16 students generated between 3 to 7 different explanations for each multiple-choice option. This showed a wide diversity to students' ideas. Yet, at the same time, students' thinking clustered around a subset of these explanations. We split the most *common explanations* (given by 3 or more students) from the *uncommon explanations* (given by fewer than 3 students). Common explanations were, on average, given by 7.5 students (SD = 4.3 students) and accounted for 76% of all explanations generated. All common explanations will be discussed further in the results. However, our primary result depends more on the nature of students' dominant and alternative answers and how closely each aligns with the canonical physical model of projectile motion.

TABLE I. The number of unique explanations coded for each option of Two Boats Q1 and Q2.

|  | # of explanations (# of common explanations given by 3 or more students) | | |
|---|---|---|---|
|  | A | B | C |
| Two Boats Q1 | 3 (1) | 5 (2) | 6 (2) |
| Two Boats Q2 | 5 (3) | 6 (1) | 7 (2) |

### A. Even students who choose the wrong answer can give the correct explanation

One noteworthy finding is that even when students chose the wrong answer, their alternative explanations still sometimes contained the correct reasoning.

For Two Boats Q1, the correct answer was B (Target 2 is hit first). One valid explanation noted that the vertical peak of shell 2 is lower, using this to infer that the time in the air for shell 2 was shorter. Another answer that we considered valid is that shell 2 had a more direct path and/or lower launch angle. In the case of equal initial speeds, a lower launch angle does correctly imply less time in the air. Therefore, we saw mention of a "more direct path" as indicating a productive resource that could help students understand the angle dependence of time.

Using students' dominant rating on part 3 (answer rating #2) as an indication of which answer they would choose, few students chose the correct answer option for Two Boats Q1. Only 1 student believed choice B was the most likely answer, and 2 students considered it tied with other explanations as the most likely to be correct. Of these 3 students, one gave the "lower peak height" explanation and one gave the "more direct path" explanation.

The majority of students believed C (hit at the same time) to be the most likely answer for Two Boats Q1. 12/16 students believed it was the most likely answer, and 1 student considered it tied with another option. One question we aimed to answer with our new approach was whether these students could generate the correct explanations when explicitly prompted for alternatives. Of the 13 students who believed strongly in choice C, 11 provided a valid explanation for why B might be the answer. 5 students gave the "lower peak height" explanation, 4 students gave the "more direct path" explanation, and 2 students gave both of these explanations. Although standard assessment approaches (which include multiple-choice tests) would capture these students' belief in the wrong answer, it would not capture their ability to generate and consider the valid explanations for the correct answer.

For Two Boats Q2, the majority of students chose incorrect option A: target 1 is hit first (9 students chose it as most likely, 2 tied). Yet, when generating explanations for answer C (the correct answer), three of those students gave the correct explanation: because the shells reach the same peak height, they hit at the same time. Though fewer students did so, this again shows that even students who chose the wrong answer can generate the correct explanation when asked for alternatives.

### B. Many incorrect explanations contained elements of correct physical reasoning

Another finding was that even invalid explanations could have productive elements of correct physical reasoning. Kinematically, time is related to distance and speed. Many explanations correctly considered how one of these two factors related to time, but ignored the other. For example, some explanations considered distance only:

- Q1/Q2: A – Target 1 is hit first, because it is closer to the battleship (Q1: 14 students, Q2: 12 students)
- Q1: C – Both are hit at the same time, because they travel roughly the same path length. The difference is that shell 1 travels farther vertically and shell 2 travels farther horizontally. (8 students)

Although both of these explanations are incorrect, they reflect correct dependences of distance on time if the speeds were equal throughout. However, they are not. For the first explanation, shell 1 has a smaller horizontal component of velocity than shell 2. For the second explanation, the speeds of the two shells do not remain equal across their trajectories. An additional problem with the second explanation here is that one cannot assume the distances traveled are exactly equal. Yet, these incorrect explanations indicate a valid physical dependence that, used properly, can play a role in learning physics.

Similarly, some explanations considered speed only:

- Q1: C – Both are hit at the same time, because they are launched with the same initial speed. (4 students)
- Q2: A – Shell 1 hits first, if it has a greater initial speed (3 students)
- Q2: A – Shell 1 reaches the peak sooner and gravity pulls it down faster, because of the greater angle (4 students)
- Q2: B – Shell 2 hits first, because it's traveling faster to get there (12 students)

These speed only explanations rely on a common *faster means less time* resource, which is consistent with kinematics. What makes these explanations incomplete is that they do not consider the effect of distance on time. For the Q2 explanations, some incorrect assumptions are also made, such as presuming that shell 1 has a greater initial speed or that shell 1 travels faster to the peak.

A complete description of motion here will integrate distance and speed to draw valid conclusions about time (as well as consistently break down the motion into horizontal and vertical components). However, even when both distance and time are included, the conclusion may not be valid. For example, on Q2, 6 students provided this explanation for choice C: shell 2 is traveling further, but it also travels faster, so it hits at the same time as shell 1. This explanation illustrates that, along with consideration of the appropriate quantities, students also need to learn how to draw valid inferences.

### C. Constructing alternatives that aren't believed

Although this methodology can uncover alternative explanations available to students, many of these explanations were associated with low levels of belief. This is an artifact of the prompt, which asks students to come up with an explanation for each choice, even if they did not believe that explanation themselves.

The benefit of this approach is that students may reveal the correct conceptual thinking in their alternative explanations. Again, in the case of Two Boats Q1, 11 students generated the correct explanation for the correct answer, B, even though they believed that C was an equally or more likely answer.

At the same time, it may also reveal "false difficulties." To illustrate, for Two Boats Q1, none of the 14 students who gave the explanation for A "target 1 is hit first, because it is closer to the battleship" believed A to be the most likely answer. The average part 3 certainty rating of choice A across these students was only 18%. Although students provided an incorrect explanation, none of them believed the answer it presented. Similarly, for Q2, 12 students provided an incorrect explanation for choice B: Shell 2 hits first, because it's traveling faster to get there, but only 3 of these students believed B could be the correct answer. The average part 3 rating of choice B across these students was 19%.

### IV. DISCUSSION

While student thinking on a multiple-choice physics question is often categorized as correct or incorrect, asking students to provide plausible explanations for each multiple-choice option reveals that student conceptual knowledge can occupy an intermediate state: a student could provide an incorrect answer, but be able to generate the correct explanation when prompted for alternatives. To uncover these findings, we used a novel approach that combines two existing methodologies for studying conceptual knowledge: clinical interviews and multiple-choice questions.

Our findings suggest that instructional philosophies viewing introduction of the correct reasoning as a necessary and sufficient aim of teaching may be misguided. If students can generate the correct answer and elements of correct physical reasoning when asked for possible alternatives, instruction may instead need to focus on helping students construct and select the correct reasoning among (more) plausible alternatives. Consistent with this finding, some instructional approaches are designed to tap these alternative resources for learning physics, making canonical physics continuous with one's existing ideas [9,10]. These approaches may be successful, because they help students see these existing ideas as physically valid. One future direction is to examine whether our prompt to provide a justification for each multiple-choice option could enhance learning by helping students access these productive ideas.

Another future area for development is on understanding the productive features of students' alternative explanations. While experts can compare students' alternative ideas to canonical physics in an ad-hoc manner, it would be better to have a systematic framework for understanding how student reasoning aligns with (and doesn't align with) a correct physical understanding. One plausible candidate for such a framework is the formalism of causal network, which can represent the complete set of causal relations between factors in a physical situation. Because the reasoning used for qualitative physics questions is fundamentally causal rather than explicitly computational, causal networks may be well-aligned with the conceptual reasoning used on such problems. We are currently exploring the use of causal networks for describing how students' explanations can become aligned with valid causal reasoning in physics [11].

When it comes to solving physics problems, students bring multiple resources to the table. The approach in this paper presents an effective way to uncover and explore those resources. It can inform not just what students are likely to answer, but also reveal the productive alternatives that may easily come to mind.

[1] L. C. McDermott, Phys. Today **37**, 24 (1984).
[2] J. P. Smith, A. A. diSessa, and J. Roschelle, J. Learn. Sci. **3**, 115 (1993).
[3] D. Hammer, A. Elby, R. E. Scherr, and E. F. Redish, in *Transfer of Learning from a Modern Multidisciplinary Perspective*, edited by J. Mestre (Information Age Publishing, Greenwich, CT, 2005), pp. 89–120.
[4] A. A. diSessa, Cogn. Instr. **10**, 105 (1993).
[5] A. A. diSessa and B. L. Sherin, Int. J. Sci. Educ. **20**, 1155 (1998).
[6] B. L. Sherin, M. Krakowski, and V. R. Lee, J. Res. Sci. Teach. **49**, 166 (2012).
[7] L. Barth-Cohen, Instr. Sci. **46**, 681 (2018).
[8] G. Gladding, M. Selen, and T. Stelzer, *SmartPhysics* (MacMillan Education. https://www.smartphysics.com/Account/LogOn, 2015).
[9] J. Clement, D. E. Brown, and A. Zietsman, Int. J. Sci. Educ. **11**, 554 (1989).
[10] D. Hammer and A. Elby, J. Learn. Sci. **12**, 53 (2003).
[11] E. Kuo, N. K. Weinlader, B. M. Rottman, and T. J. Nokes-Malach (Under Review).